\documentclass{ws-ijmpd}
\usepackage[super,compress]{cite}
\begin{document}


%
\catchline{}{}{}{}{}
%

\title{NEW MECHANISM OF ACCELERATION OF PARTICLES BY STELLAR BLACK HOLES
}

\author{OSMANOV ZAZA
}

\address{School of Physics, Free University of Tbilisi\\
Tbilisi, 0159, Georgia
\\
z.osmanov@freeuni.edu.ge}





\maketitle

\begin{history}
\received{Day Month Year}
\revised{Day Month Year}
\end{history}

\begin{abstract}
In this paper we study efficiency of particle acceleration in the
magnetospheres of stellar mass black holes. For this purpose we
consider the linearized set of the Euler equation, continuity
equation and Poisson equation respectively. After introducing the
varying relativistic centrifugal force, we show that the charge
separation undergoes the parametric instability, leading to
generation of centrifugally excited Langmuir waves. It is shown that
these waves, via the Langmuir collapse damp by means of the Landau
damping, as a result energy transfers to particles accelerating them
to energies of the order of $10^{16}$eV.
\end{abstract}

\keywords{acceleration of particles; cosmic rays; black holes}

\ccode{PACS numbers: Need to be added!}


\section{Introduction}
Recently we have published an article, where we have developed a
novel model of particle acceleration via the damping of
centrifugally excited Langmuir waves (Ref. \refcite{zev}). In
particular, we have considered magnetospheric particles close to
active galactic nuclei (AGNs) and found that the particle
acceleration is composed of several steps: in the first stage the
electrostatic waves are efficiently excited pumping rotational
energy of the magnetosphere, on the next stage the already amplified
Langmuir waves transfer their energy to plasma particles,
accelerating them by means of the "Langmuir collapse" (henceforth:
Langmuir-Landau-Centrifugal Drive (LLCD)). The mentioned mechanism
strongly depends on the relativistic effects of rotating
magnetospheres, which are present not only in AGNs but also in the
galactic sources. In particular, in (Refs. \refcite{screp2,screp1})
we have considered the Crab-like and newly born millisecond pulsars
and it has been shown that the LLCD can lead to extremely high
energy electrons.

The centrifugally excited electrostatic waves has been considered
for pulsars (Ref. \refcite{incr}) and AGN (Ref. \refcite{incr3})
where it has been found that by means of the varying centrifugal
force charge separation becomes parametrically unstable, creating
amplified Langmuir waves.

In general, it is worth noting that this double stage mechanism
requires pre acceleration of particles, that can be provided by a
direct centrifugal acceleration (Refs. \refcite{osm7}). It is worth
noting that relativistic centrifugal force is a crucial physical
element in astrophysical objects endowed with rotating
magnetospheres, energy density of magnetic field exceeds that of the
plasma. The plasma particles are, thus, forced to follow the field
lines, which in turn are corotating. Under these circumstances the
particles moving along the magnetic field lines, undergo powerful
centrifugal acceleration.

This process is extremely efficient close to the light cylinder
surface (a hypothetical zone, where the linear velocity of rotation
exactly equals the speed of light), where energy losses become
strong as well. In particular, the accelerated particles encounter
soft photons always present in the magnetospheric media. These
photons via the inverse Compton scattering (ICS) might saturate
energy gain driven by the centrifugal effects. Another mechanism
that could potentially limit the maximum attainable energies is the
so-called breakdown of the bead on the wire (BBW) approximation.
During the motion along the field lines, charged particles also
gyrate around them and the dynamics can be studied in the framework
of the bead on the wire approximation. Particles are bound by the
Lorentz force, which is not enough to keep particles close to the
field lines, when particles approach the light cylinder. This
process leads to deviation of the particles from the field lines
terminating the subsequent acceleration.

Although, by means of the direct centrifugal acceleration the
particles gain relatively low relativistic energies, they are high
enough for subsequent acceleration to ultra high energies via the
LLCD.

Apart from AGNs and pulsars, there is another class of astrophysical
objects, which are characterized by effects of rotation: stellar
black holes (SBH). It is strongly believed that SBHs are formed by
the gravitational collapse of a massive star and are members of
$X$-ray binary systems (Refs. \refcite{bhbinary}). The rotating
magnetosphere is formed by the falling accreting matter, which
transfers from a companion star. On the other hand, taking into
account the number of stars, which potentially might produce such
black holes, it is estimated that there are as many as ten million
to a billion SBHs in the Milky Way alone. Therefore, it is
interesting to apply the new mechanism of acceleration to the SBHs
and study efficiency of a corresponding process and the role of
these objects in producing the high energy cosmic rays.

In this article we apply the mechanism of LLCD to the SBHs,
analyzing the results versus physical parameters. The paper is
organized as follows: In Section 2 we introduce the mechanism of
LLCD applying it to SBHs and obtain results and in Sect. 3 we
summarize them.

\section{Theoretical model} \label{sec:main}
%
%
%
%

In this section we outline the theoretical model of LLCD examining
the centrifugal excitation of electrostatic waves, the subsequent
development of the Langmuir collapse, leading to the Landau damping
of the generated waves.

By taking into account the accretion rate onto the black hole (Ref.
\refcite{shapiro})
\begin{equation}
\label{accr} \dot{M}\approx 1.2\times 10^{12}\times
M^2_{{10}}\times\left(\frac{T_{\infty}}{10^4K}\right)^{-3/2}g\;s^{-1},
\end{equation}
where $M_{{10}}\equiv M/10M_{\odot}$ is the normalized mass of the
SBH, $M_{\odot}\approx 2\times 10^{33}$g is the Solar mass and
$T_{\infty}$ is the temperature of the accreting matter far from the
black hole, one can show that the bolometric luminosity, $L = \eta
\dot{M}c^2$, is given by
\begin{equation}
\label{lumin} L\approx 1.1\times 10^{32}\times
M^2_{{10}}\times\frac{\eta}{0.1}
\times\left(\frac{T_{\infty}}{10^4K}\right)^{-3/2}erg\; s^{-1}.
\end{equation}

The proposed mechanism of LLCD strongly depends on relativistic
effects of rotation, dynamically provided by strong magnetic fields,
therefore, it is important to estimate the angular velocity of the
black hole
\begin{equation}
\label{rotat} \Omega\approx\frac{a c^3}{GM}\approx
10^{3}\frac{a}{M_{10}}rad\;s^{-2},
\end{equation}
and the equipartition magnetic field (when the magnetic energy
density is of the order of the emission energy density) in the light
cylinder area (Ref. \refcite{osm7},
\begin{equation}
\label{mag} B\approx\sqrt{\frac{2L}{R_{lc}^2c}}\approx 2.9\times
10^3\times\frac{a}{0.1}\times\left(\frac{\eta}{0.1}\right)^{1/2}
\times\left(\frac{T_{\infty}}{10^4K}\right)^{-3/4}\;G,
\end{equation}
where $0<a\leq 1$ and $R_{lc}=c/\Omega$ is the light cylinder
radius. One can straightforwardly check that the gyroradius of a
proton with the Lorentz factor, $10$ is much less than the kinematic
lengthscale, $R_{lc}$, which in turn means that the particles will
co-rotate with the magnetic field lines.

The magnetospheric plasma consists of protons and electrons and
therefore, their dynamics is crucial for studying the LLCD. For
describing the centrifugally excited Langmuir waves we apply the
$1+1$ approach (Ref. \refcite{membran}) following the model
developed in (Ref. \refcite{zev}) and linearize the system of
equations, composed of the Euler equation
 \begin{equation}
\label{eul3} \frac{\partial p_{\beta}}{\partial
t}+ik\upsilon_{\beta0}p_{\beta}=
\upsilon_{\beta0}\Omega^2r_{\beta}p_{\beta}+\frac{e_{\beta}}{m_{\beta}}E,
\end{equation}
the continuity equation
\begin{equation}
\label{cont1} \frac{\partial n_{\beta}}{\partial
t}+ik\upsilon_{{\beta}0}n_{\beta}, +
ikn_{{\beta}0}\upsilon_{\beta}=0
\end{equation}
and the Poisson equation
\begin{equation}
\label{pois1} ikE=4\pi\sum_{\beta}n_{\beta0}e_{\beta},
\end{equation}
where by ${\beta}$ we denote the species index of stream particles:
electrons and protons, $k$ is the wave number, $\upsilon_{0\beta}(t)
\approx c\cos\left(\Omega t + \phi_{\beta}\right)$ is the zeroth
order velocity and $r_\beta(t) \approx
\frac{c}{\Omega}\sin\left(\Omega t + \phi_{\beta}\right)$ is the
radial coordinate (Ref. \refcite{zev}), $c$ is the speed of light,
$\phi_{\beta}$ is the initial phase of particles, $p_{_{\beta}}$ is
the first order dimensionless momentum ($p_{_{\beta}}\rightarrow
p_{_{\beta}}/m_{_{\beta}}$), $m_{_{\beta}}$ and $e_{\beta}$ are the
particle's mass and charge respectively, $E$ is the induced
electrostatic field and $n_{_{\beta}}$ and $n_{_{\beta0}}$ are the
perturbed and unperturbed Fourier components of the density. It is
worth noting that the aforementioned behaviour of the radial
coordinate (and velocity as well), is physically applicable until
the particle reaches the light cylinder surface and the dynamics
cannot be extended for timescales exceeding $P/4$.

After applying the following anzatz
\begin{equation}
\label{ansatz}
n_{\beta}=N_{\beta}e^{-\frac{iV_{\beta}k}{\Omega}\sin\left(\Omega t
+ \phi_{\beta}\right)}
\end{equation}
to the aforementioned equations they reduce to
\begin{equation}
\label{ME1} \frac{d^2N_p}{dt^2}+{\omega_p}^2 N_p= -{\omega_p}^2 N_e
e^{i \chi},
\end{equation}
\begin{equation}
\label{ME2} \frac{d^2N_e}{dt^2}+{\omega_e}^2 N_e= -{\omega_e}^2 N_p
e^{-i \chi},
\end{equation}
where $\omega_{e,p}\equiv\sqrt{4\pi
e^2n_{e,p}/m_{e,p}\gamma_{e,p}^3}$ and $\gamma_{e,p}$ are the
relativistic plasma frequency and the Lorentz factor for the stream
particles, $\chi = b\cos\left(\Omega t+\phi_{+}\right)$, $b =
\frac{2ck}{\Omega}\sin\phi_{-}$, $2\phi_{\pm}= \phi_p\pm\phi_e$ and
$J_{\mu}(x)$ is the Bessel function of the first kind. By applying
the Fourier transform, Eqs. (\ref{ME1}-\ref{ME2}) lead to the
dispersion relation (Ref. \refcite{zev})
\begin{equation}
\label{disp} \omega^2 -\omega_e^2 - \omega_p^2  J_0^2(b)= \omega_p^2
\sum_{\mu} J_{\mu}^{2}(b) \frac{\omega^2}{(\omega-\mu\Omega)^2}.
\end{equation}

It is clear from Eq. (\ref{disp}) that the system undergoes the
parametric instability for the following resonance condition
$\omega_r = \mu_{res}\Omega$. To define the growth rate we follow
the standard method discussed in (Ref. \refcite{screp1}) and express
the frequency as $\omega = \omega_r+\Delta$. After substituting it
into Eq. (\ref{disp}) it reduces to
\begin{equation}
\label{disp1} \Delta^3=\frac{\omega_r {\omega_p}^2
{J_{\mu_{res}}(b)}^2}{2},
\end{equation}
with the imaginary part of the solution for $\Delta$, which in turn
gives the increment of the instability (Ref. \refcite{zev})
\begin{equation}
 \label{grow}
 \Gamma= \frac{\sqrt3}{2}\left (\frac{\omega_e {\omega_p}^2}{2}\right)^{\frac{1}{3}}
 {J_{\mu_{res}}(b)}^{\frac{2}{3}},
\end{equation}
where $\mu_{res} = \omega_e/\Omega$.

\section{Discussion} \label{sec:res}
%
%
%
%

We have already discussed in the introduction that the LLCD requires
pre-acceleration, which can be guaranteed by direct centrifugal
mechanism, strongly limited by two major factors: the ICS and the
BBW approximation. As it has been shown in (Ref. \refcite{osm7}) the
corresponding Lorentz factors are given by
\begin{equation}
\label{gic} \gamma_{_{IC}}\approx\left(\frac{6\pi mc^4}{\sigma_T
L\Omega}\right)^2,\;\;\;\;
\gamma_{_{BBW}}\approx\frac{1}{c}\left(\frac{e^2L}{2m}\right)^{1/3}
\end{equation}
where $m$ is the particle's mass and $\sigma_T\approx 6.65\times
10^{-25}$cm$^{-2}$ is the Thomson cross-section. One can
straightforwardly check that for both: electrons and protons the BBW
approximation is the dominant factor limiting the maximum attainable
energies and the expression for the corresponding maximum Lorentz
factor writes as
\begin{equation}
\label{gme} \gamma_{max}\approx 800\times
M^{2/3}_{{10}}\times\left(\frac{m}{m_e}\right)^{-1/3}
\times\left(\frac{\eta}{0.1}\right)^{1/3}
\times\left(\frac{T_{\infty}}{10^4K}\right)^{-1/2}.
\end{equation}
Therefore, for the reasonable physical parameters the electrons
might reach the Lorentz factor of the order of $800$ and the protons
- respectively $65$, which as we will see is quite enough for the
LLCD to work efficiently.

\begin{figure}
  \resizebox{\hsize}{!}{\includegraphics[angle=0]{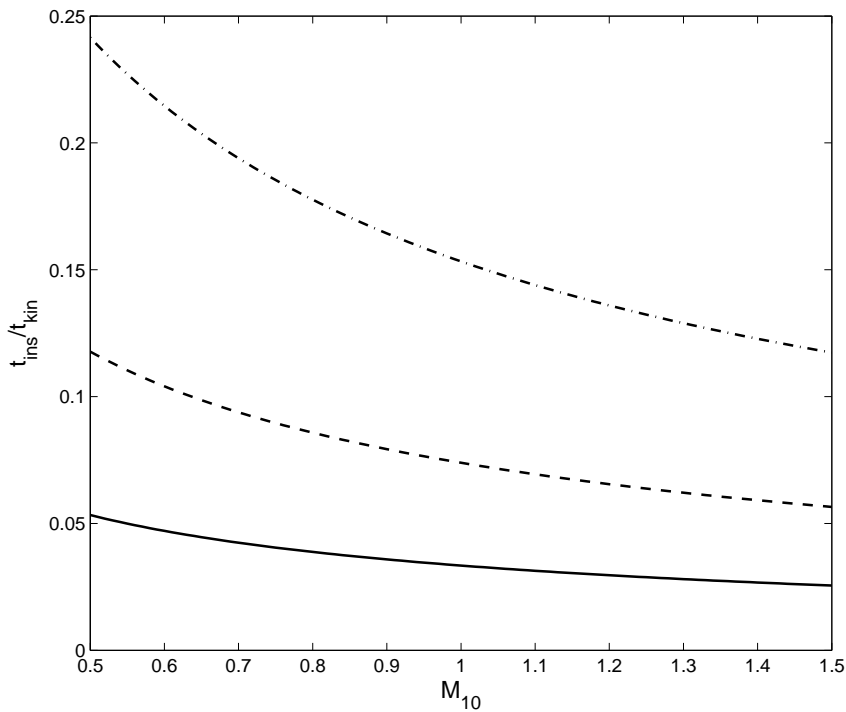}}
  \caption{Dependence of $t_{ins}/t_{kin}$ on the
  SBH mass. The set of parameters is $\gamma_e = 800$,
$\gamma_p = 1.5$ (solid line), $\gamma_p = 2.5$ (dashed line),
$\gamma_p = 4$ (dashed-dotted line), $\eta = 0.1$, $a = 0.1$,
$T_{\infty} = 10^4$K.}\label{fig1}
\end{figure}
We have discussed in the introduction that inside the magnetosphere
(inside the light cylinder surface) plasma particles are in the
frozen-in condition, and therefore co-rotate with magnetic field
lines. In particular, it is straightforward to check that the Larmor
radius of protons for the aforementioned Lorentz factors is by four
orders of magnitude less than $R_{lc}$. This means that the
magnetosphere is rotation powered and therefore, the longitudinal
(along the magnetic field lines) electric field is neutralized by
charge distribution leading to the density, $n_0$, of the order of
$\sim\Omega B/(2\pi ec)$ also for Black hole magnetospheres (Refs.
\refcite{zev},\refcite{neronahar}). As an example we consider a
stream with relativistic electrons, $\gamma_e\sim 800$, examining
three different cases of protons $\gamma_p = 1.5; 2.5; 4$. On Fig.
\ref{fig1} we display the behavior of $\frac{t_{ins}}{t_{kin}}$
versus $M_{10}$, where $t_{ins}\sim 1/\Gamma$ is the instability
timescale and $t_{kin} = 2\pi/\Omega$ is the kinematic timescale.
The set of parameters is $\gamma_e = 800$, $\gamma_p = 1.5$ (solid
line), $\gamma_p = 2.5$ (dashed line), $\gamma_p = 4$ (dashed-dotted
line), $\eta = 0.1$, $a = 0.1$, $T_{\infty} = 10^4$K. We show the
plot versus the black hole mass in the range $(10-15)\times
M_{\odot}$. It is worth noting that up to now the largest known
(extragalactic) stellar black hole has mass of the order of
$15.6\times M_{\odot}$ (Ref. \refcite{bulik}). As it is clear from
the figure, for the aforementioned parameters $t_{ins}<t_{kin}$,
which means that the centrifugally induced electrostatic instability
is quite efficient.

An effect of the second stage - the damping of electrostatic waves -
can be even stronger if the "Langmuir collapse" precedes it. In
particular, by means of the high frequency pressure the particles
are pushed out from the perturbation area creating the so-called
caverns: relatively small density regions. This in turn, leads to
the acceleration of plasmons towards the caverns, leading to the
modulation instability (Ref. \refcite{zakharov}).

Assuming the quasi neutrality of the plasma one can show that in a
simplified $1$D-case, the process of the "Langmuir collapse" in the
rest frame of the fluid is described by the following set of
equations (Ref. \refcite{zakharov})
\begin{equation}
\label{ez1} \left[\frac{\partial^2}{\partial
t^2}-3\lambda_D^2\omega_p^2\frac{\partial^2}{\partial
x^2}-\omega_p^2\right]E = \frac{\delta n}{n_0}\omega_p^2E,
\end{equation}
\begin{equation}
\label{ez2} \left[\frac{\partial^2}{\partial
t^2}-\lambda_D^2\omega_p^2\frac{\partial^2}{\partial
x^2}\right]\delta n=\frac{1}{16\pi m_p}\frac{\partial^2E^2}{\partial
x^2},
\end{equation}
where $\lambda_D\equiv \sqrt{k_BT_e/(4\pi n_0e^2)}$ is the Debye
length scale, $k_B$ is the Boltzman constant, $T_e$ is the electron
temperature, $n_0$ is the unperturbed ion number density, $\delta n$
is the electron number density perturbation and $E$ is the induced
electric field.

Induced electric field is very significant for studying the
development of the Langmuir collapse. For this purpose it is worth
noting that perturbation corresponding to density perturbation is
small compared with the unperturbed density and therefore frequency
of plasmons and correspondingly energy should be constant (Ref.
\refcite{arcimovich})
\begin{equation}
\label{E2a} \int d^q{\bf r}\mid E\mid^2 = const,
\end{equation}
where $q$ denotes spacial dimensions, which might equal $1,2,3$
depending on a concrete physical system. From the above equation it
is evident that the electric field behaves with distance as
\begin{equation}
\label{E2} \mid E\mid^2\propto\frac{1}{l^q}.
\end{equation}
As we have discussed in (Ref. \refcite{zev}) inside the light
cylinder surface, where the magnetic field dominates plasmas, the
particles are forced to move along the field lines, therefore such a
physical system is strongly defined by $1D$ dynamics leading to the
following behavior of the high frequency pressure $P_{hf}\propto
1/l$, where we have taken into account the relation $P_{hf}\propto
E^2$ (Ref. \refcite{zakharov,arcimovich}). On the other hand, the
plasmons inside the caverns have kinetic and potential energies of
the same order of magnitude
\begin{equation}
\label{k} k^2\lambda_D^2\sim\frac{\mid\delta n\mid}{n_0},
\end{equation}
leading to the following behaviour of the thermal pressure
$P_{th}=k_BT\delta n\propto 1/l^2$. Therefore, for smaller
lengthscales the thermal pressure overcomes the high frequency
pressure and the collapse cannot develop. This situation drastically
changes outside the magnetosphere (light cylinder surface), where
particles do not follow the field lines any more and become free.
Under such conditions particle dynamics is described by three
dimensional geometry and consequently it is evident that unlike the
previous case, now, the high frequency pressure dominates leading to
the development of the efficient collapse. By applying
Eqs.(\ref{k},\ref{ez2}) and neglecting the term corresponding to the
thermal pressure one obtains
\begin{equation}
\label{dnt} \frac{\partial^2\delta n}{\partial
t^2}\approx\frac{\delta n\mid E\mid^2}{16\pi nm_p\lambda_D^2},
\end{equation}
which after taking into account the following relations $\mid
E\mid^2\sim 1/l^3$ and $\delta n\sim 1/l^2$, leads to (Ref.
\refcite{zakharov})
\begin{equation}
\label{E2} \mid E\mid^2\approx \mid
E_0\mid^2\left(\frac{t_0}{t_0-t}\right)^2
\end{equation}
\begin{equation}
\label{l} l\approx l_0\left(\frac{t_0}{t_0-t}\right)^{-2/3},
\end{equation}
where $t_0$ denotes the moment of the collapse. As we see, this
instability has an explosive character, which is more efficient than
standard exponentially amplified processes.

\begin{figure}
  \resizebox{\hsize}{!}{\includegraphics[angle=0]{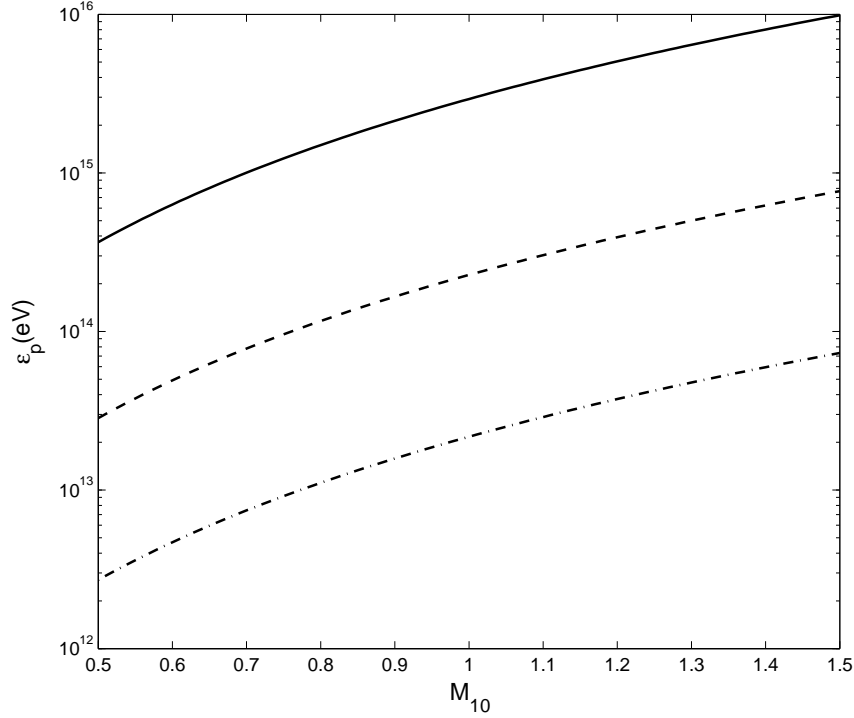}}
  \caption{Behaviour of $\epsilon_p\left(eV\right)$  versus the
  SBH mass. The set of parameters is $\gamma_e = 800$,
$\gamma_p = 1.5$ (solid line), $\gamma_p = 2.5$ (dashed line),
$\gamma_p = 4$ (dashed-dotted line), $\eta = 0.1$, $a = 0.1$, $f =
0.01$, $T_{\infty} = 10^4$K.}\label{fig2}
\end{figure}
In (Ref. \refcite{zev}) we have found that in the light cylinder
area energy gain of a proton is given by
\begin{equation}
\label{energy} \epsilon_p\approx\frac{E^2}{8\pi n}.
\end{equation}
The electric field writes as (Ref. \refcite{zev})
\begin{equation}
\label{elec} E\approx 4\pi en_0\Delta r\times\frac{\Delta
r^{3/2}}{l_c^{3/2}},
\end{equation}
where $\Delta r = R_{lc}/(2\gamma_p)$ is the thickness of a thin
layer close to the light cylinder, $l_c\approx c/(\omega_0\sqrt{f})$
is the lengthscale of the cavern, $\omega_0=\sqrt{4\pi ne^2/m}$ and
$f = \delta n/n_0$. The number density of protons outside the
magnetosphere, $n$, is not governed by rotation any more and can be
approximated by assuming the spherical symmetry of accretion
\begin{equation}
\label{n} n=\frac{L}{4\eta\pi m_pc^2\upsilon R_{lc}^2}.
\end{equation}
Here $\upsilon = \sqrt{{GM}/R_{lc}}$ is the velocity of an infalling
matter.

By considering the beam electrons with $\gamma_e = 800$ and assuming
$\eta = a = 0.1$, $T_{\infty} = 10^4$K the final energy of protons
writes as (see Eq. \ref{energy})
\begin{equation}
\label{energy1} \epsilon_p\left(eV\right)\approx 7\times
10^{14}\times
M_{10}^{3}\times\left(\frac{f}{10^{-2}}\right)^3\times\left(\frac{2}{\gamma_p}\right)^5.
\end{equation}
On Fig. \ref{fig2} the dependence of $\epsilon_p\left(eV\right)$ on
$M_{10}$ is whown. The set of parameters is $\gamma_e = 800$,
$\gamma_p = 1.5$ (solid line), $\gamma_p = 2.5$ (dashed line),
$\gamma_p = 4$ (dashed-dotted line), $\eta = 0.1$, $a = 0.1$, $f =
0.01$, $T_{\infty} = 10^4$K. As it is clear from the plots, the
energy of protons is a continuously increasing function of the SBH
mass (see Eq. (\ref{energy1})) and we see that the SBHs might
provide ultra high energy protons, $\sim 10^{16}$eV, via the LLCD.
Although, for more massive SBHs the result will be even higher.

In general, the accelerating particles might undergo strong energy
losses, therefore, the mechanisms limiting the acceleration process
has to be taken into account. We have already discussed in the
introduction that by means of the synchrotron mechanism the
particles transit to the ground Landau states very soon and
therefore it does not influence the acceleration of particles any
more.

The ultra-high energy protons interact with soft photons, which are
always present in a medium around SBHs, therefore, the corresponding
scattering has to be taken into account. In (Ref. \refcite{blum}) it
has been found that for extremely energetic particles, the mentioned
interaction operates in the so-called Klein-Nishina regime leading
to the cooling timescale, which is proportional to the proton energy
(Ref. \refcite{or09}). This in turn means that the inverse Compton
mechanism does not impose a significant constraint on energy gain.

Apart from the above mentioned processes there is another mechanism
- curvature radiation - which potentially might be responsible for
limiting the maximum attainable energies of protons. By taking into
account the corresponding cooling timescale,
$t_{cur}=\epsilon_p/P_{cur}$, where $P_{cur} =
2e^2\epsilon_p^4/(3m_p^4c^3\rho)$ is the single particle radiation
power and $\rho\sim R_{lc}$ is the curvature radius of a trajectory,
one can straightforwardly show that for the typical parameters of a
medium surrounding the SBHs (see Fig. \ref{fig1}, Fig. \ref{fig2})
the curvature timescale exceeds that of the Landau collapse
instability (Ref. \refcite{zev,arcimovich})
\begin{equation}
\label{incrcol} t_{_{LC}}\approx\gamma_p\left[\frac{\langle
E^2\rangle e^2m_p}{4k_BT}\right]^{-1/2},
\end{equation}
by many orders of magnitude, which means that the curvature
radiation is also negligible and does not impose any constraints on
achievable energies.

As we have already mentioned in the introduction, the population of
SBHs in the Milky Way might be at least ten million, which means
that they might be very significant in studying the high energy
cosmic rays.

\section{Summary} \label{sec:summary}
%
%
%
%

\begin{enumerate}

     \item The main novelty of the manuscript is that the recently studied new
     mechanism of particle acceleration - LLCD process we applied to
     SBHs. The present mechanism is composed of two major stages. In the
     first stage centrifugal force parametrically excites the unstable
     electrostatic waves, efficiently extracting energy from
     rotation. On the second stage the Langmuir collapse develops
     significantly enhancing the energy gain via the Landau
     damping.

     \item By linearizing the set of equations composed of the Euler
     equation, continuity equation and Poisson equation we have
     shown that by means of the relativistic centrifugal force the
     very unstable Langmuir waves are generated. It has been shown
     that the instability timescale is less than the kinematic
     timescale, indicating high efficiency of the energy pumping
     process into the waves. As a next step we have considered the
     possibility of Langmuir collapse and it has been found that
     energy transfer from the waves into the particles is so
     efficient that neither the inverse Compton, nor the curvature
     emission affect particle acceleration. As a result, it is shown
     that by means of the LLCD the protons in the medium surrounding the
     SBH with mass exceeding the solar mass 15 times might reach energies of
     the order of $10^{16}$eV.

 \end{enumerate}

The aim of the paper was to show a role of rotation in the jet-like
structure of the Crab pulsar. The study was only focused on the
dynamic behaviour of particles, moving along prescribed (in the RF)
co-rotating channels.

An important restriction in the present model is the consideration
of a single particle approach, whereas it is clear that in a general
case dynamics of particles is strongly influenced by collective
phenomena. Therefore, it would be interesting to explore the dynamics of
magnetocentrifugally accelerated particles in this context.

The formalism developed in Section 2 is valid for both Schwarzschild
and Kerr black holes even though in the subsequent analysis we
completely neglected the gravitational effects and considered only a
special-relativistic case. One of the tasks of further study will be
to check how magnetocentrifugal acceleration along prescribed
trajectories works in fully relativistic situations and physically
realistic astrophysical scenarios.

\section*{Acknowledgments}

The research was partially supported by the Shota Rustaveli National
Science Foundation grant (N31/49).

\end{document}